\documentclass[aps,prd,showpacs,twocolumn]{revtex4}

\usepackage{latexsym}
\usepackage{amsfonts,amssymb}
\usepackage{graphicx,epsfig}
\usepackage{psfrag}

\newcommand{\be}{\begin{equation}}
\newcommand{\ee}{\end{equation}}
\def\bq{\begin{eqnarray}}
\def\eq{\end{eqnarray}}
\def\n{\nonumber}
\def\t{\tau}
\def\ti{\tilde}

\def\a{\alpha}
\def\th{\theta}
\def\G{\Gamma}

\def\s{\sigma}

\def\La{\Lambda}
\def\Laf{\Lambda}
\def\ra{\rightarrow}
\def\lra{\Longrightarrow}

\begin{document}

\title{Localized gravity on FRW branes}

\author{Parampreet Singh\footnote{e-mail: param@iucaa.ernet.in} and Naresh Dadhich\footnote{e-mail: nkd@iucaa.ernet.in }}
\address{Inter-University Centre for Astronomy and Astrophysics,\\
Post Bag 4, Ganeshkhind, Pune-411 007, INDIA.}


\begin{abstract}                                    
We study the system of Schwarzschild anti de Sitter (S-AdS) bulk and FRW brane 
for localization of gravity; i.e. zero mass gravitons having ground state on 
the brane, and thereby recovering the Einstein gravity with high energy 
correction. It has been known that gravity is not localized on AdS brane with 
AdS bulk. We prove the general result that gravity is not localized 
for dynamic branes whenever 
$\Lambda_4 < 0$, and is localized for the curvature index $k = 1$ only when 
$\Lambda_4 > 0$ and black hole mass $M \neq 0$, else it is localized for all 
other FRW models. If the localization is taken as the brane world compatibility
 criterion for cosmological models, then it would predict that negative 
cosmological constant on the brane is not sustainable. 
\end{abstract} 

 
\maketitle       

\section{Introduction}

The idea that our universe has dimensions more than four has been around
since the first attempts to unify fundamental forces. The Kaluza-Klein theories
were one of the first attempts towards this direction in which the size 
of extra 
dimension was taken to be of the order of Planck length. In its new avatar in the form
of brane world scenarios, it has attracted much attention in recent times.
In these models our physical universe is envisioned as a four dimensional
hypersurface in a five dimensional bulk spacetime. The standard model
matter is confined to the brane but gravity, by its universal character,
can propagate in the extra dimension. 

Early attempts in this direction were
based on extra dimensions being of the order of millimeter \cite{arkani} which is
less than the current observational limits on low scale gravity. 
The warped extra dimension models of Randall and Sundrum (RS) \cite{rs1,rs2} have seen  
intense activity. Their two brane model had a brane with negative tension which led to anti 
gravity \cite{shiromizu} and was hence
 ruled out. In their single brane model, the flat brane sits in an AdS bulk
which is $Z_2$ symmetric. The $Z_2$ symmetry of the extra dimension arises from the 
reduction of M theory to $E_8 \times E_8$ heterotic
 string theory \cite{witten}. RS were able to  recover
Newton's inverse square law with correction terms. These correction terms arise from the 
massive Kaluza-Klein(KK)
 modes on the brane and their effect can be tested at sub-millimeter level \cite{hoyle}.
There have also come about quite a few generalizations,  for eg.  in the form of thick branes 
(see for eg. \cite{csaki}), AdS branes \cite{karch} and brane 
models without $Z_2$ symmetry \cite{colyuri}. 

The brane and bulk spacetimes are joined through the Israel junction
conditions \cite{israel}. The Einstein equation on the brane is
modified and there are various interesting cosmological consequences of
the modifications \cite{cosmo1}, including CFT effects \cite{cft}. Some of the important 
solutions
of bulk spacetime are branes in AdS bulk and FRW branes in Schwarzschild-AdS (S-AdS) bulk 
\cite{karch, bdel, kraus, s-ads, sd2}.

Extra dimensions, though introduced to deal with problems in standard
model of particle physics and also motivated by unification schemes, give a new color to 
the cosmological aspects
of our physical universe. The S-AdS bulk modifies the Friedmann equation on the brane which 
now contains two extra terms. First is a matter density squared term and the other is a dark 
radiation term propotional to
the mass parameter of the bulk black hole. The latter is contribution of the bulk Weyl 
curvature. The density squared term dies out
at low energies and the dark radiation term is severely constrained by the
big bang nucleosynthesis. Ultimately, at low energies there survives no
observational signature which can distinguish this theory from the 
standard cosmology. However, there have been some attempts by incorporating
string theory motivated correction terms in the action which result in
modifications available  at the low energy sector \cite{gb}. They can
be tested against current observations, eg. supernova Ia observations \cite{sn}. 
Such observations have provided a 
valuable tool to test standard cosmological model and other novel ideas
like those with variable cosmological constant \cite{vishwa}. In one of the
studies based on AdS-CFT correction to S-AdS bulk it was found that
the theory provides a definite low energy signature which fits very well
with the supernova Ia observations for a wide range of parameters and also brings relief
to the age of the universe problem in the standard cosmological model \cite{svd}.

However, before one addresses the cosmological problems it is important
to answer the question of localization of gravity on the brane harboring such
models. Localization is not guaranteed a priori as we shall see that it does not happen 
for all  FRW models, and in particular for negative cosmological constant on the brane. 
Apart from $\Lambda$, the curvature index $k=0, \pm1$ and presence of black hole in the bulk 
which will make Weyl curvature in the bulk non-zero will also have bearing on this question. 
In particular, for the conformally non flat Nariai bulk, it has been shown that there can not 
exist any normalizable bound massless graviton \cite{sd1}. 
Another example of non localization is provided by AdS brane in AdS bulk \cite{karch} and 
we shall in this investigation establish that this feature holds good more generally 
even when brane is non empty. We shall establish the general result: \emph{gravity is never 
localized whenever $\Laf < 0$ on a dynamic brane, is localized for $k = 1$ only when $\Laf > 0, M \neq 0$ and is 
else always localized on the brane for all other FRW models}.

The cosmologically interesting FRW brane resides in S-AdS bulk for which
the most pertinent question of localization was addressed by authors for a
special case \cite{sd2}. Here we would expand upon our previous work and deal with all 
possible cosmological scenarios. We would consider the perturbations of the bulk metric to 
study allowed ground state of massless graviton on the brane together with motion of brane 
in the bulk spacetime. It is well known that a small off tuning from
the critical brane tension makes the brane dynamic \cite{kraus}. For an FRW expanding 
universe on the brane, brane will have to be moving in bulk spacetime. Hence the brane 
equation of motion would also play a critical role alongside the perturbation equation in 
determining localization of gravity on the brane. In particular, localization would require brane 
to be ever expanding which would rule out negative cosmological constant on the brane. This is a 
definite prediction of the brane world cosmology. The most critical and attractive feature
of brane world gravity is that it is localized on the brane. Localization could hence 
serve as compatibility criterion for cosmological models. Further, it is interesting that brane 
dynamics in the bulk, apart from other physical considerations, also determines that mass of 
the bulk black hole must be positive. 

In the previous studies, the Gaussian normal coordinates
for AdS bulk were employed, we shall however carry out the perturbation analysis 
in the natural curvature coordinates for S-AdS bulk spacetime. This should not matter in the 
end for localization must be coordinate independent physical property. In Sec. II, 
we carry out the perturbation analysis and recover RS model results in Sec. III. In the 
following three sections are discussed the cases of the curvature parameter $k = 0, 1 $ and
$ -1$ for the vanishing and non vanishing mass of the bulk black hole. We would show that 
a number of interesting cosmological models are indeed compatible and allowed, in 
particular inflationary universes with positive
cosmological constant on the brane which is favored by the current
observations \cite{sn}. For completeness we shall also consider static brane which will not be of 
practical use for it can not harbor expansion. This will be followed by conclusion.
   
In the next section we study the graviton fluctuations of the bulk metric
and derive the wave equation governing the effect of extra dimension on its
modes. 

\section{Perturbation of the bulk metric} 
In this section we would derive the wave equation of the metric fluctuations  
of the S-AdS bulk, 
\be
d s^2 = - e^{2 \beta} \, d t^2 + e^{- 2 \beta} d y^2 +  y^2 \bigg[ \frac{d r^2}{1 - k r^2} 
+ r^2 d \Omega_2^2 \bigg] \label{eq:metric}
\ee
where 
\be
e^{2 \beta} = \left( k + \frac{y^2}{l^2} - \frac{M}{y^2} \right).
\ee
Here $k = +1, 0, -1$ is the spatial curvature index, $l$ is the 
curvature radius of the bulk spacetime and $M$ is the mass parameter of the black hole. It  
comes as a constant of integration in the solution of the Einstein equation in the bulk,
\be
G_{ab} = - \Lambda_5 \, g_{ab}, \hspace{1cm} \Lambda_5 = - 6/l^2.
\ee
The Latin indices run from 0...4 and the Greek indices would run from 0...3. 
 The event horizon of the black hole would be at
\be
y_h^2 = \frac{l^2}{2} \, \left(- k + \sqrt{k^2 + \frac{4 M}{l^2}} \right)
\label{eq:horizon}
 \ee
which thus restricts $M \geq 0$ for $k = 0, 1$ and $M \geq - l^2/4$ for $k = -1$. The extra 
dimension is the radial coordinate of the bulk black hole.
 The perturbations of the above metric  over
the background metric $g_{ab}^{(B)}$, i.e. $h_{ab} = g_{ab} - g_{ab}^{(B)}$, satisfies the following wave
equation 
\be
\Box_{5} h_{ab} = \Lambda_5 \, h_{ab}.
\ee
We are interested in finding out the effect of extra dimension on the
behavior of gravitons on the brane. We choose  the metric perturbations in the extra dimension to vanish,
\be
h_{t \, y} = h_{r \, y} = h_{\th \, y} = h_{\phi \, y} = h_{y \, y} = 0 
\ee
and further impose the transverse-traceless gauge conditions on the
metric fluctuations,
\be
\nabla^{\mu} h_{\mu \nu} = 0, \hspace{1cm} h^{\mu}{}_{\mu} = 0.
\ee
With these conditions  the wave equation can be written down as
\be
\Box_{5} h_{ab} = A^i_{ab,i} + \Gamma^i_{ij} \, A^j_{ab} - \Gamma^i_{aj} A^j_{ib} -\Gamma^i_{bj} A^j_{ia} = \Lambda_5\, h_{ab}
\ee
where
\be
A^i_{ab} := g^{ij}\left( h_{ab,j} - \G^k_{aj} h_{kb} - \G^k_{bj} h_{ak} \right).
\ee
We would look for  the plane wave solutions of the form
\be
h_{a b} = e_{ab} \, \Psi(y) e^{i \, p_{j} \, x^{j}}
\ee
where the polarization tensor $e_{ab}$ is orthogonal to a unit
time-like vector $u^a = (1,0,0,0,0)$ i.e. $e_{ab} \, u^b = 0$.
Solving the wave equation and using $p^2 = - m^2$ as the constant of separation of variables it can be shown that $\Psi(y)$ satisfies the following equation,
\be
\left(\frac{y^2}{l^2} - \frac{M}{y^2} + k \right) \Psi'' + \left(\frac{3 M}{y^3} - \frac{k}{y} + \frac{y}{l^2} \right) \Psi' - \frac{2 M}{y^4} \Psi + m^2 \Psi = 0  \label{eq:y}
\ee
where prime denotes a derivative with respect to $y$. This equation can be written down in the form 
\be
\Psi'' + a_1(y) \Psi' + a_2(y) \Psi = 0
\ee
which can be transformed into a wave equation form with the transformation $\Psi(y) = \phi(y) \psi(y)$
where $\phi(y)$ is such that it eliminates the first order derivative terms. Substituting this in eq.(\ref{eq:y}) we get the
desired wave equation,
\be
- \frac{1}{2} \psi(y)'' + V \psi(y) = m^2 \psi(y). \label{eq:yeq}
\ee
This equation holds all the information about the effect of extra dimension on
the gravitons on the brane. The form of $V$ is quite complicated. The potential
is made up of two parts,  $V = V_f + V_m$, where the free part is  
\bq
V_f =  &-& \n
 \frac{6 l^2 (5 M y^4 - k y^6)}{8 (y^5 + l^2(- M y + k y^3))^2} 
+ \frac{M l^2}{y^2 (y^4 - M l^2 + k l^2 y^2)} \\
 &-&  \frac{y^8 + l^4( - 15 M^2 + 22 k M y^2 - 3 k^2 y^4)}{8 (y^5 + l^2(- M y + k y^3))^2}
\eq
and the interaction part is given by
\be
V_m = - \frac{m^2 l^2 y^2 - 2 m^2(y^4 - M l^2 + k l^2 y^2)}{2(y^4 - M l^2 + k l^2 y^2)} .
\ee  
It is composed of two types, one is free of mass
of graviton and the other depends on it. So different gravitons
would see a slightly different potential produced by the bulk. Such potentials
are analog to the ones obtained in oscillator problems where natural frequency
is a function of coordinates. Since our brane is curved and 
dynamic, such interaction terms are to be expected. In the following sections
we would discuss how  different values of $k$ and $M$ affect propagation of graviton modes.

The effective four dimensional
cosmological constant on the brane and the Gravitational constant
can be determined from the Israel junction conditions \cite{israel}
\be
\Lambda = \frac{\Lambda_5}{2} + \frac{16 \pi^2}{3} \, G_5^2 \sigma^2, \, 
G = \frac{4 \pi}{3} \, G_5^2 \, \s
\ee
where $\sigma$ is the brane tension and $\La$ and $G$ are the four dimensional
effective cosmological and gravitational constants. 
The value of effective cosmological constant is hence determined by the 
bulk parameters and the brane tension. By a proper fine tuning of the brane
tension the effective cosmological constant on the brane can be made to vanish.
We would denote this special fine tuned brane tension by $\s_0$ which equals
$(3/4 \pi G_5 l)$. Hence for $\s\lesseqqgtr \s_0$ we have  
$\Lambda\lesseqqgtr 0$.

The junction conditions also determine how the brane moves in the bulk. 
In RS case the brane was
static but in general the brane would be moving and that
would affect the boundedness of the modes. The junction conditions also reflect
that the FRW brane evolves in time and its trajectory is determined by
the following equation \cite{kraus},
\be
{\dot y}^2 - \frac{8 \pi G}{3} \, \rho \,\left(1 + \frac{\rho}{2 \s}
\right) y^2 - \frac{\Lambda}{3} \, y^2 -  \frac{M}
{y^2}  + k = 0. \label{eq:bdynawm}
\ee
where 
\be
\Lambda = - \frac{3}{l^2}  \, (1 - \s^2/\s_0^2) .
\ee
Here  dot refers to the derivative w.r.t. proper time ($\t$).

The brane trajectory hence also depends on the presence of matter on the
brane. The normalization of the ground state wavefunction is determined by the shape of the 
potential $V$. For massless graviton to remain bound to the brane, the potential $V_f$  should remain 
negative for the entire range of $y$ once it crosses the horizon, as
exhibited in Figs. 1 and 2.  Further, for the massive modes $V_m > 0$ at large $y$. 
The massless mode can be viewed as $ m \ra  0$ limit and  
hence for normalization of the ground state wavefunction $V_f$ should
asymptote to zero.
  This would certainly be so for expanding brane and 
it could be made so for a static brane by proper choice of parameters. On the other hand, it would not 
be so for an oscillating brane.

Localization of gravity on the brane would thus hinge on two properties, one $V_f < 0$ 
for the entire range of $y$ and two the brane is ever expanding and has no bounce. The 
latter requirement means $\dot y$ must never vanish. We shall now analyze for what choice of 
the parameters this condition could be satisfied? Putting ${\dot y} = 0$ in eq.(\ref{eq:bdynawm}) 
and neglecting the density squared term at low energies $(\rho \ll \s)$ we get,
\be
\Laf \, y^4 - 3 k \, y^2 + 8 \pi \, G \, \mu \, y + 3 M = 0 \label{eq:roots} 
\ee
where $\rho_ = \mu / y^3$ for a matter dominated universe, which would be the case at low energy.
Clearly for $\Laf < 0$, this equation would always admit a real positive root indicating oscillatory 
motion for the brane. Thus even in the presence of black hole on the bulk and matter on the brane and 
irrespective of the curvature index $k$, no localization of massless graviton can occur  on the brane for negative cosmological 
constant. This is the generalization of the AdS non-localization result \cite{karch}. Similar would be the 
case for $k = 1$ and $\Laf = 0$. The case $k = 1$ and $\Laf > 0$ would have either
two or none positive real roots. It would hence be possible to make suitable choice of parameters to 
avoid bounce. For all other cosmological cases, there would exist expanding brane solution. We shall 
rule out $\Laf < 0$ and $k =1, \Laf = 0$ cases from further considerations.

At large $y$, energy 
density term in  eq.(\ref{eq:bdynawm}) will not be very effective for it will at best go as 
$y^{-3}$ or $y^{-4}$ according as matter or radiation dominance. Thus asymptotically 
brane motion would essentially be governed by the cosmological constant $\Lambda$. We 
would hence for our study set $\rho = 0$, in which case
eq.(\ref{eq:bdynawm}) becomes
\be
{\dot y}^2 - \frac{\Lambda}{3}  y^2 -  \frac{M}{y^2}  + k = 0
\label{eq:bdyna} .
\ee 
Since this equation is non linear in its highest order of derivative, it will
not have unique solution. The metric on the brane would be FRW,
\be
d s_4^2 = - d\t^2 +  y^2(\t) \left(\frac{d r^2}{1 - k r^2} + d \Omega_2^2
\right)
\ee   
where  $d\t^2 = exp(2 \beta(t))(1 - exp(- 4 \beta(t)) (d y/ d t)^2 ) d
t^2 $. Thus a physically acceptable solution of eq.(\ref{eq:bdyna}) with a normalizable
bound massless mode of the Schroedinger equation, eq.(\ref{eq:yeq}), would ensure 
localization of gravity on the brane harboring FRW cosmologies. Our physical universe 
could then be envisioned as emerging out of the horizon like a white hole \cite{wh}.
 
\section{Recovering the Randall-Sundrum result}

In the Randall-Sundrum case, bulk is AdS with a flat brane, i.e. the
effective cosmological constant on the brane is zero. This case arises from
eq.(\ref{eq:metric}) for $k = M = 0, \s = \s_0$, hence, 
the metric becomes
\be
d s^2 = -\frac{y^2}{l^2} d t^2 + \frac{l^2}{y^2} d y^2 + y^2 \left(d r^2 + r^2 d r^2 + r^2 \, 
\sin^2\th d \phi^2 \right) .  
\ee
This metric can be transformed into the RS form by 
using the following transformation
\be
\eta = l \, \ln(l/y), \hspace{1cm} \lra \hspace{1cm} d \eta = - \frac{l}{y} dy \label{eq:RStrans}
\ee
Defining $d {\ti x}_1^{2} + d {\ti x}_2^{2} + d {\ti x}_3^{2} \,  \equiv \, l^2 (d x_1^2 + d x_2^2 + d x_3^2) \, = \, l^2 (
d r^2 + r^2 d r^2 + r^2 \, \sin^2\th d \phi^2)$ we get
\be
d s^2 = d \eta^2 + e^{- 2 \eta/ l} (- d {\ti t}^{2} + d {\ti x}_1^{2} + d {\ti
x}_2^{2} + d {\ti x}_3^{2}) \ee
where $d {\ti t} \, \equiv \,  d t$. From the brane trajectory equation  
we get
\be
{\dot y}^2 = 0, \hspace{1cm} \lra \hspace{1cm} y(\t) = constant,
\ee 
i.e. the brane is static in the bulk.  
In the RS case the brane was at $\eta = 0$, this  implies through
our transformation (eq.(\ref{eq:RStrans})) that the brane in our case is at $y = l$. Hence 
the constant above equals $l$. 

The bulk potential $V$ is as shown in Fig. 1. The Schroedinger equation, eq.(\ref{eq:yeq}), in
this case reduces to 
\be
 \frac{1}{2} \psi(y)'' + \frac{1}{8 y^2} \psi(y) + \frac{m^2 l^2}{2 y^2} \psi(y) = 0 .\label{eq:rsweq}
\ee

\begin{figure}[tbh!]
\epsfig{figure=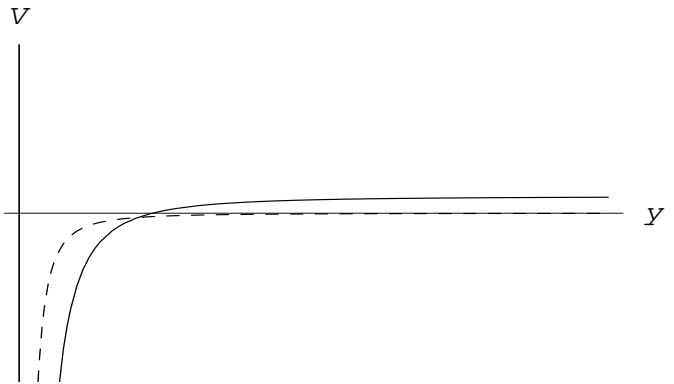,height=2in,width=3.5in,angle=0}
{\caption{\small Behavior of the potential for $k = 0, M = 0$. The dashed 
and solid lines respectively indicate massless and massive graviton modes. RS brane is 
located at $y = l$.}}\end{figure}

The wavefunction near the brane is then the linear combination of Bessel functions,
\be
\psi(y) = \sqrt{y} \, \left( C_1 \, J_0(m y) + C_2 \, Y_0(m y) \right) 
\ee
which at the brane ($y = l$) becomes
\be
\psi(y) = \sqrt{l} \, \left(C_1 \, J_0(m l) + C_2 \, Y_0(m l) \right) .
\ee
 From Fig. 1 we note that 
 for the massless mode $V \ra 0$ as $y \ra \infty$.  For all other
massive modes, $V > 0$ as $y \ra \infty$. 
 The behavior
of the potential tells that in addition to the continuum of massive modes there
is a possibility of the existence of discrete massive modes for gravitons of
mass less than $1/2l$. 
 However, for very small $l$  the massive modes effectively form a 
continuum extending down
to the massless case. The difference in asymptotic behavior of the potential
for the massless and massive modes also reflects the decoupling between the
two.
The boundary conditions on the wavefunction on the brane and far away from it lead to $C_1 = m^{1/2}$ and $C_2 = \pi \sqrt{m}$ thus yielding $\psi_m = (m l)^{1/2}$ on
the brane as in the RS case. At large distance from the brane the
continuum wavefunction tends to that of plane waves. 

The massive modes will produce  corrections over the Newtonian gravity 
reproduced by the normalizable massless graviton. The massive modes would
constitute a Yukawa like interaction between two mass point particles
$m_1$ and $m_2$ on the brane with an appropriate plane wave density measure.
The effective gravitational potential on the 
brane becomes,
\bq
U({\ti r}) &\sim &  \n G \frac{m_1 m_2}{\ti r} + G \, l \, \int_{0}^{\infty} dm \frac{m_1 m_2}{\ti r} \, e^{- m {\ti r}} m l \\ 
U(\ti r) &\sim &  G \frac{m_1 m_2}{\ti r} \left(1 + \frac{l^2}{\ti r^2} \right),
\eq
where ${\ti r}^2 = {\ti x_1}^2 + {\ti x_2}^2 + {\ti x_3}^2$.
The contributions from the massive modes are exponentially suppressed for $m$ greater than $1/{\ti r}$ and hence one
recovers Newtonian gravity on the brane.

\section{Other cosmological models for $k = 0$} 

We would now study other cosmological cases which arise apart
from RS with $k = 0$. The simplest are the ones in which the mass of
the bulk black hole is zero, but brane tension is not fine tuned to $\s_0$. 
If the brane 
tension is tuned less than the critical value $\s_0$ 
there would exist no physical brane trajectory, however for tuning it more
than $\s_0$ the brane would expand forever in the bulk.

Case (I): \, $\s > \s_0$ ,  $M = 0$. In this case $\Laf > 0$ and
hence eq.(\ref{eq:bdyna}) becomes,
\be
{\dot y}^2 - \frac{\Laf}{3}  \, y^2 = 0, \hspace{0.5cm} \lra \hspace{1cm} y(\t) = l \,
e^{\sqrt{\Laf/3} \tau}.
\ee
 The brane starts
 expanding out exponentially from its initial position and keeps expanding forever. 

The Schroedinger equation for this case is the same as for the RS case. The only difference 
is that it is now moving and does not stay fixed. Since the brane is now moving we would parameterize the small deviations from
 the initial position by a dimensionless parameter defined as
 $\alpha = 1 - y_i^2/y^2$, where $y_i$ is the initial position of the brane i.e. $y_i = l$. Solving eq.(\ref{eq:rsweq}), the wavefunctions for this case are,
\be
\psi(y) = \sqrt{y} \left(C_1 \, J_0( \sqrt{1 - \a } m y) + C_2 \, Y_0(\sqrt{1 - \a } m y) \right) .
\ee 
The boundary conditions imply that $C_1 = m^{1/2}$ and $C_2 = \pi m^{1/2}$. 
The potential plot is again shown by Fig. 1. The continuum as before is suppressed by a 
factor of $m l$ and the  massless mode as before yields effective Newtonian gravity
near the horizon. As the brane expands out, the discrete modes also turn to
continuum.
The Hubble parameter for this case is given by
\be
H = \frac{{\dot y}}{y} = \sqrt{\frac{\La}{3}}
\ee
and thus an inflationary universe is born.  
We thus see that when the brane tension is greater than
the RS case (i.e. $\Lambda > 0$) we recover Newtonian gravity with
correction due to massive KK modes, and when it is less than the RS case
(i.e. $\Lambda < 0$) there does not exist a physically permissible motion for the brane.

Now we would consider the cases when the mass parameter of the bulk black hole
is non zero. There would occur the horizon at $y_h = (\sqrt{M} l)^{1/2}$. The graviton potential
is shown in Fig. 2.  We first consider the case of $\Laf = 0$.

Case (II): \, $\s = \s_0$ ($\Laf = 0$), $M \neq 0$. This case is the RS case with a
black hole in the bulk. The motion of the brane is determined by,
\be
{\dot y}^2 - \frac{M}{y^2}  = 0 .
\ee
Using the initial condition $y(0) = 0$ we get
\be
y(\t) = \pm \, \sqrt{2}  \, M^{1/4} \, \sqrt{\t}, 
\ee
the physically acceptable solution. The brane evolves to expand as $y(\t) \sim \t^{1/2}$. 
The mass of the black hole tends to
increase the rate of expansion of the brane. 

\begin{figure}[tbh!]
\epsfig{figure=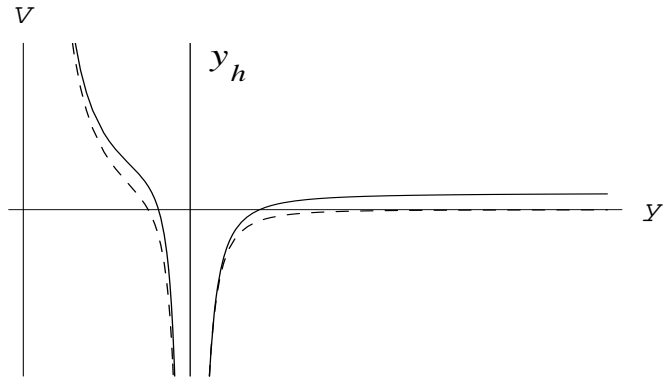,height=2in,width=3.5in,angle=0}
{\caption{\small Potential plot for $k = 0, M > 0$. The horizon occurs at
$y_h = \sqrt{M} l$. For the massless mode it asymptotes to zero.}}
\end{figure} 

Case(III): \,  $\s > \s_0$ , $M \neq 0$. In this case $\Laf > 0$
and the dynamics of the brane is given by 
\be
{\dot y}^2 - \frac{\Laf}{3} \,  y^2 -  \frac{M}{y^2}  = 0 . 
\ee
which leads to the following physical solution consistent with $Z_2$-symmetry,
\be
y(\t) = \left(\sqrt{\frac{3 M}{\Laf}}  \, \sinh(2 \sqrt{\Laf/3} \t) \right)^{1/2} .
\ee
The brane expands out exponentially and
keeps expanding.
The potential for massless and massive modes is shown in Fig. 2 by
dashed and solid curve respectively. To the right of the event horizon it resembles 
RS potential. 

We now solve for the Schroedinger equation (eq.(\ref{eq:yeq})) in near the horizon approximation by
introducing a   dimensionless parameter $\a = 1 - \sqrt{M} l/y^2$. The wavefunctions are,
\be
\psi(y) = \sqrt{y} \left(C_1\, I_{-\gamma} (\nu \, y^2) + C_2 \, I_{\gamma} (\nu\, y^2) \right)
\ee
where
\be
\gamma = \frac{\sqrt{5 \a^2 - 4}}{4 \a}, \, ~ \nu = \sqrt{(2 \a^2 + \a - 1)/(2 \a M)}\, \frac{m}{2}.
\ee
The boundary conditions on the  wavefunction near and far away from the brane
lead to $C_1 =  m^{\gamma + 1/2} l^{\gamma} M^{-1/4}$ and $C_2 = 0$. The massive modes are $m l$ suppressed on the brane and as the brane expands out in the negative potential the discrete modes join the continuum. The presence
of discrete modes has also been linked to temperature of the brane \cite{temp}.
The trapped massless mode leads to recovery of Newtonian potential 
with massive modes contributing to the RS correction.
 As the brane expands
and the discrete modes enter continuum, the temperature of the brane decreases
which is in accordance with the fact that the universe cools down as it 
expands.

The Hubble parameter for the case $\s = \s_0$ is given by
\be
H = \frac{{\dot y}}{y} = \frac{1}{2 \tau}.
\ee
The Einstein's equations on the
brane are 
\be
G^{\t}_{\t} = - \Laf - 8 \pi G \, {\ti \rho}, 
\ee
and
\be
G^{r}_{r} = G^{\th}_{\th} = G^{\phi}_{\phi} =  - \Laf + 8 \pi G \, {\ti p}
\ee
where ${\ti \rho}$ and ${\ti p}$, and which are given by
\be
{\ti \rho} = \frac{3\, M }{8 \pi G \, y^4}, ~~ ~~ {\ti p} = \frac{M}{8 \pi G \, y^4}, \label{eq:dp}
\ee
satisfying the radiation equation of state ${\ti \rho} = 3 {\ti p}$.
 For the case $\s > \s_0$ the Hubble parameter takes the form 
\be
H = \frac{{\dot y}}{y} = \sqrt{\frac{\La}{3}} \, \coth(2 \sqrt{\La/3}\, \tau)
\ee
with the same effective energy density and pressure. In the
cases when $M$ is zero, ${\ti \rho} = {\ti p} = 0$. Thus we see
that for curvature index $k = 0$, except for $\Laf < 0$ all other FRW models 
are allowed. 

We would now consider the models for $k = 1$.

\section{Cosmological models for $k = 1$}  

In this case when $k = 1$ and if there is no black hole in the bulk i.e. $M = 0$,
there can be no localization of gravity possible. This is because $V > 0$, always, which would imply 
absence of bound state on the brane and hence ruled out from further consideration. This 
shows that presence of black hole is necessary to sustain FRW cosmology for the positive curvature 
brane. 

However the situation changes dramatically with the presence of black hole for which the  
horizon would occur at $y_h^2 = l^2 (-1 + \sqrt{1 + (4 M/l^2)})/2$. The potential is similar to the one shown in Fig. 1, 
with a negative well at the horizon and $V_f$ asymptoting to zero. We would now turn to 
motion of the brane.  For $k = 1$, since the cases $\Laf \leq 0 $ have already been ruled out, we are thus left 
with the solitary case of \, $\s > \s_0$ ,  $M \neq 0$. In this case the
effective cosmological constant on the brane is positive. This case has been 
reported \cite{sd2} earlier. The brane trajectory is given by 
\be
{\dot y}^2 - \frac{\Laf}{3} \, y^2 - \frac{M}{y^2} + 1 = 0 ,
\ee
which has the following solution
\be
y(\t) = \sqrt{\frac{3}{2 \La}} \, \bigg[ 1 + n \sinh(2 x) - \cosh(2 x)
\bigg]^{1/2} 
\ee
where $n = 2 \sqrt{M \Laf/3}$ and $x = \sqrt{\Laf/3} \tau$.
For $n < 1$ the brane expands till $\tanh(2x) = n$ and then recollapses
to origin. Thus localization of massless mode is not possible for $n < 1$.
However, for $n > 1$ solution is inflationary and thus the brane would expand out
exponentially. It could take off from the origin and then crosses the horizon 
and expands forever. With the expansion of the brane
the discrete massive modes turn into continuum. To obtain the wavefunctions
we solve the Schroedinger equation in the approximation $M \ll l^2$. The
horizon in this approximation occurs at $y_h = \sqrt{M}$. We parameterize
the location off the brane by introducing a parameter $\a = 1 - M/y^2 $ and
solving the Schroedinger equation we get,
\be
\psi(y) =  \sqrt{y} \left( C_1 I_{-\gamma/2} (\nu \, y) + C_2 
I_{\gamma/2} (\nu \,  y) \right)
\ee   
where
\be
\gamma = \sqrt{1 - 4 l^4(1 - 2 \a)(M - 2 \a l^2)/M^3}
\ee
and
\be
\nu=\sqrt{(1 - \a )(M - 2 \a l^2)} m l /M .
\ee
The boundary conditions on the wavefunction at the brane and far away from it lead to $C_1 = (m l)^{(1 + \gamma)/2} M^{\gamma/4}$ with $\gamma < 2$ and $C_2
= 0$. The massive continuum is $m l$ suppressed at the location of the brane.
The massless mode generates the Newtonian potential and the high energy correction due to massive modes dies out as the brane expands out from the horizon.
The Hubble parameter for this case is,
\be
H = \frac{\dot y}{y} = \sqrt{\frac{\Laf}{3}} \, \frac{\bigg[ n \,
\cosh(2 x) - \sinh(2 x) \bigg]}{\bigg[ 1 + n \, \sinh(2 x) - \cosh(2 x)
\bigg]} \, \,  
\ee
which implies an inflationary universe at large $\t$. The effective energy density 
and pressure on the brane would once again be the same as given earlier in eq.(\ref{eq:dp}).

In this case, note that $n > 1$ is essential for localization which would mean $M > 3/(4 \Laf)$. That puts a 
constraint on the black hole mass.
 
\section{Cosmological models for $k = -1$}

In this case the potentials for $M = 0$ and $M \neq 0$ are similar to the
one shown in Fig. 2. In the absence of the black hole the horizon lies at
$y_h = l$. This case is similar to the flat $k = 0$ case. 

Case(I):  \, $\s = \s_0$, $ M = 0$. This case is analog of RS
case for $k = -1$. The brane dynamics equation yields,
\be
{\dot y}^2 - 1 = 0, ~~ ~~ \lra ~~ ~~  y(\t) = \pm \t .
\ee
The motion is hence unbounded. The brane expands out linearly from the origin, crosses the horizon and continues expanding forever. 

Case (II): \, $\s > \s_0$ , $M = 0$. In this case  eq.(\ref{eq:bdyna}) takes the form
\be
{\dot y}^2 - \frac{\Laf}{3} \,  y^2 - 1 = 0, ~~ \lra ~~ y(\t) = \frac{\sinh(\sqrt{\Laf/3} \t)}{\sqrt{\Laf/3}} .
\ee
Thus we have an inflationary solution. The brane
 expands out exponentially from the origin and 
expands for ever.

Introducing a parameter $\a = 1 - l^2/y^2$ we solve the  Schroedinger equation to get the following solution,
\be
\psi(y) = \sqrt{y} \left( C_1 J_{-\gamma}(\nu y) + C_2 J_{\gamma}(\nu y) \right)
\ee
where $\gamma = \sqrt{\a^2 - 4}/(2 \a)$ and $\nu = 2 m /\sqrt{\a}$ with 
constants  $C_1 = m^{1/2 - \gamma} l^{-\gamma}$ and $C_2 = m^{1/2 + \gamma} l^{\gamma}$. The Hubble parameter for the case $\s = \s_0$ brane is
\be
H = \frac{{\dot y}}{y} = \frac{1}{\tau}
\ee
whereas for the $\s > \s_0$ brane it is given by
\be
H = \frac{{\dot y}}{y} = \sqrt{\frac{\La}{3}} \, \coth(\sqrt{\La/3} \, \tau) .
\ee
It is easy to check using Einstein's equations on the brane that the
effective energy density and pressure on the brane vanish.

We now consider the cases when there is a black hole in the bulk. 
The event horizon is at  $y_h^2 = l^2 (1 + \sqrt{1 + (4 M/l^2)})/2$.

Case(III): \,   $\s = \s_0$ , $ M \neq 0$. The brane dynamics is
given by,
\be
{\dot y}^2 -  \frac{M}{y^2} - 1 = 0
\ee
which results in two physical solutions
\be
y_1(\t) =  \sqrt{\t (\t + 2\sqrt{ M})}, ~~ y_2(\t) =  \sqrt{\t (\t - 2\sqrt{ M})} . \label{eq:mneg1}
\ee
For the first solution the brane starts expanding out from the origin at 
$\t = 0$ and keeps expanding, whereas in the second case it
does not start expanding out from the origin until $\t = 2 \sqrt{M}$.
The first solution has a time hole in the region $0 \geq \tau \geq - 2 \sqrt{M}$ whereas the second has 
in the region $0 \leq \tau \leq 2 \sqrt{M}$.
For large $\t$ the behavior of the brane is $y \propto |\t|$.

Case(IV): \,  $\s > \s_0$ , $\Laf > 0$, $ M \neq 0$. The trajectory of the 
brane in this case is governed by
\be
{\dot y}^2 - \frac{\Laf}{3} \,  y^2 -  \frac{M}{y^2} - 1 = 0.
\ee
This equation has two physical solutions,
\be
y_1(\tau) =  \, \sqrt{\frac{3}{2 \La}} \, \bigg[\cosh(2 x) + n \, \sinh(2 x) - 1 \bigg]^{1/2} \label{eq:mneg2}
\ee
and
\be
y_2(\tau) =  \, \sqrt{\frac{3}{2 \La}} \, \bigg[\cosh(2 x) - n \, \sinh(2 x) - 1 \bigg]^{1/2} . \label{eq:mneg3}
\ee
The brane would be ever expanding for  the first and for $n < 1$ for the second solution. 
That means there would always exist an expanding solution. 

Note that though for $k = -1$
case a black hole with negative value of mass parameter is allowed through
the horizon equation with the constraint $M \geq -l^2/4$, it is
clear from eqs.(\ref{eq:mneg1},\ref{eq:mneg2},\ref{eq:mneg3}) that a negative
$ M $ would make the solutions imaginary. Thus we come to an important result that
the black hole mass parameter which enters in the metric as  constant
of integration and can in principle be positive or negative, cannot be 
negative.

\begin{table*}[tbh!]                                                           \caption{Summary of localization of gravity for various FRW brane models. The case $k=0$, $M=0$ and
$\Laf = 0$ is the static RS case. }
\begin{ruledtabular}
\begin{tabular}{|cccc|}
\hline
 ~~ ~~ $k$       & $M$  & $\Laf$ & Localization of Gravity   \\
\hline
\hline
~~ ~~ 0 &  0 & $<$ 0 & $\times$ \\
~~ ~~ 0 & 0 & 0 & $\surd$ \\
~~ ~~ 0 & 0 & $>$ 0 & $\surd$ \\
~~ ~~ 0 & $\neq$ 0 & $<$ 0 & $\times$ \\
~~  ~~ 0 & $\neq$ 0 & 0 & $\surd$ \\
~~ ~~ 0 & $\neq$ 0 & $> $0 & $\surd$ \\
\hline
~~ ~~ 1 &  0 & $<$ 0 & $\times$ \\
~~ ~~ 1 & 0 & 0 & $\times$ \\
~~ ~~ 1 & 0 & $>$ 0 & $\times$ \\
~~ ~~ 1 & $\neq$ 0 & $<$ 0 & $\times$ \\
~~  ~~ 1 & $\neq$ 0 & 0 & $\times$ \\
~~ ~~ 1 & $\neq$ 0 & $> $0 & $\surd$ \\
\hline
~~ ~~ - 1 &  0 & $<$ 0 & $\times$ \\
~~ ~~ - 1 & 0 & 0 & $\surd$ \\
~~ ~~ - 1 & 0 & $>$ 0 & $\surd$ \\
~~ ~~ - 1 & $\neq$ 0 & $<$ 0 & $\times$ \\
~~  ~~ - 1 & $\neq$ 0 & 0 & $\surd$ \\
~~ ~~ - 1 & $\neq$ 0 & $> $0 & $\surd$ \\
\hline
\end{tabular} 
\end{ruledtabular}
\end{table*}    

Solving the Schroedinger equation in the
approximation $M \ll l^2$ and introducing a parameter $\a = 1 - l^2/y^2$
the wavefunctions can be obtained as
\be
\psi(y) = \sqrt{y} \left( C_1 I_{-\gamma/2}(\nu y) + C_2 I_{\gamma/2}(\nu y) \right) 
\ee
where 
\be
\gamma = \sqrt{(\a -4)/\a}, ~~ \nu = \sqrt{5 \a M - 3 M - 2 \a l^4 m^2}/(\sqrt{2 \a} l^2) .
\ee
The boundary conditions on the brane and far away from the brane determine
$C_1 = m^{\gamma + 1/2} l^{\gamma}$ and $C_2 = 0$. The massive continuum
as before is suppressed by $m l$ factor on the  brane. The Hubble
parameters for the $\Laf = 0$ case ($y_1(\t)$ and $y_2(\t)$) are given by
\be
H_1 = \frac{{\dot y}}{y} = \frac{\tau + \sqrt{M}}{\tau(\tau + 2 \sqrt{M})}
\ee
and
\be
H_2 = \frac{{\dot y}}{y} = \frac{\tau - \sqrt{M}}{\tau(\tau - 2 \sqrt{M})}.
\ee
 For the $\Laf > 0$ case, the Hubble parameters corresponding to the two solutions read as  
\be
H_1 =  \frac{{\dot y}}{y} = \sqrt{\frac{\La_4}{3}} \, \frac{\bigg[ n \, \cosh(2 x) + \sinh(2 x) \bigg]}
{\bigg[- 1  + \cosh(2 x) + n \, \sinh(2 x)\bigg]} \, \,  
\ee
and
\be
H_2 =  \frac{{\dot y}}{y} = \sqrt{\frac{\La_4}{3}} \, \frac{\bigg[ n \, \cosh(2 x) - \sinh(2 x) \bigg]}
{\bigg[ 1  - \cosh(2 x) + n \, \sinh(2 x)\bigg]} \, \,  
\ee
with the same ${\ti \rho}$ and ${\ti p}$ as in eq.(\ref{eq:dp}). 

We have summarized the results of localization of gravity for various
possible cosmological scenraios in Table 1.

\section{Static branes}

We now wish to look for the brane solutions when the brane remains 
static in the bulk potential. Though such solutions can not harbor expansion
and hence are not of much cosmological interest. It is however interesting to see how
a proper fine tuning of parameters can keep the brane static which would
otherwise be moving. Static brane in other than S-AdS bulk has also been considered \cite{gergely}.  

Since the matter density on a static brane would remain constant, the 
$\rho$ and $\rho^2$ terms would effectively behave as $\Laf$ term in 
eq.(\ref{eq:bdynawm}), which for  ${\dot y} = 0$ becomes,
 \be
 \frac{\eta}{3} \,  y^2  + \frac{M}{y^2} - k  = 0 \label{eq:stateq}
\ee
where
\be
\eta = 8 \, \pi \, G \, \rho \,\left(1 + \frac{\rho}{2 \s}
\right)  + \Lambda .
\ee

For the brane to remain fixed at a location, time derivative of eq.(\ref{eq:stateq}) must also vanish which would 
require  $2 \sqrt{M \eta/3} = k$. Putting that back in eq.(\ref{eq:stateq}) we get  $y^2 = 2 M/k$ . 
It is interesting to see a balance between the black hole mass and energy density on the brane required to keep it fixed 
at a location which is again determined by the black hole mass. Clearly $k$ cannot be negative, however $\Laf$ on the 
static brane could be positive, negative or zero. The localization compatibitlity condition for the static brane is that 
it cannot have negative curvature.

When it is flat $k = 0$, both $M$ and $\eta$ have to vanish and we have the generalization of the RS 
model for non empty brane, and $\Laf < 0$ necessarily. The brane could be located anywhere other than 
$y = 0$. When $k$ is non zero, it must be $k = 1$, the brane would be located at $y^2 = 2M > y_h^2$ which means it would always lie 
outside the horizon. In this case, both $M$ and $\eta$ will have to be non zero. Static brane thus brings out a close relationship 
between the bulk black hole and energy distribution on the brane.

\section{Conclusion}

Localization of gravity on the brane is certainly a critical requirement for 
brane world cosmology. We have established the general result that
{\em for a dynamic FRW brane in S-AdS bulk, gravity is never localized whenever $\Lambda_4 < 0$, is 
localized for $k = 1$ only when $\Lambda_4 > 0$ and $M\neq 0$ and else is always localized 
on the brane for all other FRW models.} 

Localization hinges on two features: one, potential for zero mass graviton to be negative and 
second, brane must always be expanding or static at a location. Except for the case $k = 1$, the 
former condition holds good without any constraint on the parameters. For $k = 1$, presence of 
black hole in the bulk is necessary to make $V_f < 0$. That is, black hole in the bulk helps 
confinement of gravitons on the brane. In all other cases, the critical role is played by eq. (19), 
which determines  whether brane would have bounce or not. For ever expansion, this equation must not have 
a real positive root. It is clear that only negative $\Laf$ and positive $k$ favor bounce and consequently oscillatory 
universe, while $M$ helps expansion. Asymptotically, $\Laf$ dominates over all others and hence when it is negative, there 
would always occur a bounce and thereby no localization. Presence of black hole critically matters only in the case of 
$k = 1$, where it turns potential negative. In all other cases, its role is facilitative rather than critical. Non negative $\Laf$ and 
non positive curvature index augurs well with expansion and from the bulk they are joined constructively by mass of the black hole.
The opposing effect comes from negative $\Laf$ and positive $k$. It is expansion which is critical for localization of gravity for 
zero mass graviton should asymptotically see vanishing potential. It would
be interesting to investigate the localization of other spin fields and
new interactions which might result in this setting, as has been done for the case of RS brane \cite{bajc}.

Though static branes are not of cosmological interest, they exhibit a direct relation between mass of the bulk 
black hole and energy distribution on the brane. A static brane cannot have negative curvature. When it is flat, both black 
hole mass and effective cosmological constant $\eta$ must vanish and this gives the non emepty brane generalization of RS brane. It 
could be located freely anywhere except at $y = 0$. When $k = 1$, both black hole mass and $\eta$ must be 
non zero and its location is fixed at $y^2 = 2M$ which would always lie outside the horizon. The brane is kept static by the balance 
between black hole mass and $\eta$, which should always be positive. On the other hand $\Laf$ could 
have any sign.

Based on localization of gravity on the brane it clearly follows that brane world cosmology makes the definite 
prediction that in a dynamic universe \emph{the cosmological constant on the brane cannot be negative and the Universe 
must be ever expanding}. This is an observationally testable prediction. The present observations indicate accelerating 
expansion which augurs well with this prediction. The main message of our study is that FRW cosmologies on the brane 
with S-AdS bulk spacetime are well founded gravitationally.

We thank Luis Anchordoqui, Roy Maartens, Carlos Nunez, S. Shankaranarayanan
and R. G. Vishwakarma for various useful discussions and comments. PS is
supported by a research grant from Council 
for Scientific \& Industrial Research.


\newpage

\end{document}